\documentclass[12pt,preprint]{aastex}

\shorttitle{BD~--22$^{\circ}$5866: A low-mass ESB4}
\shortauthors{Shkolnik et al.}

\begin{document}


\title{BD~--22$^{\circ}$5866: A Low-mass Quadruple-lined Spectroscopic {\it and} Eclipsing Binary\altaffilmark{1}\\}


\author{Evgenya~Shkolnik\altaffilmark{2}}
\affil{NASA Astrobiology Institute, Institute for Astronomy, University of Hawaii at Manoa\\ 2680 Woodlawn Drive, Honolulu, HI 96822}
\email{shkolnik@ifa.hawaii.edu}

\author{Michael C. Liu\altaffilmark{3}}
\affil{Institute for Astronomy, University of Hawaii at Manoa\\ 2680 Woodlawn Drive, Honolulu, HI 96822}
\email{mliu@ifa.hawaii.edu}

\author{I. Neill Reid}
\affil{Space Telescope Science Institute, Baltimore, MD 21218}
\email{inr@stsci.edu}

\author{Leslie Hebb, Andrew C. Cameron}
\affil{School of Physics and Astronomy, University of St. Andrews, North Haugh
St Andrews, Fife
Scotland KY16 9SS}
\email{leslie.hebb@st-andrews.ac.uk, Andrew.Cameron@st-and.ac.uk}

\author{Carlos A. Torres}
\affil{Laborat\'orio Nacional de Astrof\'isica/MCT, Rua Estados Unidos 154, 37504-364 Itajub\'a, Brazil}
\email{beto@lna.br}

\and

\author{David M. Wilson}
\affil{Astrophysics Group, Keele University, Staffordshire, ST5 5BG}
\email{dw@astro.keele.ac.uk}

\altaffiltext{1}{Based on
observations collected at the W. M. Keck Observatory and the Canada-France-Hawaii Telescope.  The Keck Observatory is operated as a scientific partnership between the California Institute of
Technology, the University of California, and NASA, and was made possible by the generous
financial support of the W. M. Keck Foundation. The CFHT is operated by the National Research Council of Canada,
the Centre National de la Recherche Scientifique of France, and the University of Hawaii.}
\altaffiltext{2}{NASA Postdoctoral Fellow}
\altaffiltext{3}{Alfred P. Sloan Research Fellow}

\begin{abstract}

We report our discovery of an extremely rare, low mass, quadruple-lined spectroscopic binary BD~--22$^{\circ}$5866 (=NLTT 53279, integrated spectral type = M0~V), found during an ongoing search for the youngest M dwarfs in the solar neighborhood. From the cross-correlation function, we are able to measure relative flux levels, estimate the spectral types of the components, and set upper limits on the orbital periods and separations.  The resulting system is hierarchical composed of K7 + K7 binary and a M1 + M2 binary with semi-major axes of $a_\mathrm{A}$sin$i_\mathrm{A}$$\leq$0.06 AU and $a_\mathrm{B}$sin$i_\mathrm{B}$$\leq$0.30 AU. A subsequent search of the SuperWASP photometric database revealed that the K7 + K7 binary is eclipsing with a period of 2.21 days and at an inclination angle of 85$^{\circ}$. Within uncertainties of 5\%, the masses and radii of both components appear to be equal (0.59 M$_{\sun}$, 0.61 R$_{\sun}$). These two tightly orbiting stars ($a$ = 0.035 AU) are in synchronous rotation causing the observed excess Ca II, H$\alpha$, X-ray and UV emission.  
The fact that the system was unresolved with published adaptive optics imaging, limits the projected physical separation of the two binaries at the time of the observation to $d_\mathrm{AB}$$\lesssim$4.1 AU at the photometric distance of 51 pc. 
The maximum observed radial velocity difference between the A and B binaries limits the orbit to $a_\mathrm{AB}$sin$i_\mathrm{AB}$$\leq$6.1 AU.  As this tight configuration is difficult to reproduce with current formation models of multiple systems, we speculate that an early dynamical process reduced the size of the system such as the interaction of the two binaries with a circumquadruple disk.
Intensive photometric, spectroscopic and interferometric monitoring as well as a parallax measurement of this rare quadruple system is certainly warranted.


\end{abstract}


\section{Introduction}\label{intro}

The multiplicity of stars is an important constraint of star formation theories as most stars form as part of a binary or higher-order multiple system (Duquennoy \& Mayor 1991, Fischer \& Marcy 1992, Halbwachs et al.~2003). 
Moreover, double- (or multi-) lined spectroscopic binaries (SBs) allow precise determination of dynamical properties including the mass ratio, and if the inclination can be determined, the individual component masses. And given their common age and metallicity, SBs are very useful in calibrating stellar evolutionary models.  



Here, we report our detection of a visually unresolved quadruple-lined spectroscopic {\it and} eclipsing (ESB4) binary BD~--22$^{\circ}$5866 (=NLTT 53279; $V = 10.1$, $J_\mathrm{2MASS} = 7.54$) composed of four low-mass stars. 
As part of our search for young M dwarfs within 25 pc (Shkolnik et al.~2006), we acquire high-resolution spectra of cool dwarfs compiled by the NStars project (Reid et al.~2003, 2004) that have X-ray luminosities comparable to or greater than Pleiades members. 
We cross-correlate these spectra with radial velocity (RV) standard stars in order to determine the RVs needed to measure galactic space motion.  This process is sensitive to finding SBs, particularly those in short-period orbits, while the sample in general is biased towards tidally-locked systems whose rapid rotation produces high chromospheric and X-ray emission. BD~--22$^{\circ}$5866 was on our list of candidate young M dwarfs because of its high fractional X-ray\footnote{X-ray data is from the ROSAT All-Sky Survey Bright Source Catalogue (Vogues et al.~1999).} and UV\footnote{Near and far UV (NUV, FUV) fluxes were measured by the GALEX (GALaxy Evolution eXplorer) All-Sky Survey (Morrissey et al.~2007).} flux ($f_{X}/f_{J}$ = 1.94$\times10^{-3}$, $f_{NUV}/f_{J}$ = 1.6$\times10^{-4}$, $f_{FUV}/f_{J}$ = 3$\times10^{-5}$) and photometrically determined distance of 20.7$\pm$3.4 pc (Reid et al.~2004). 


In this paper we decompose BD~--22$^{\circ}$5866 into a tight hierarchical quadruple system consisting of an eclipsing K7 + K7 binary and a M1 + M2 binary, and discuss the implications of this extremely unusual multiple system for the dynamical evolution of binary stars.  
BD~--22$^{\circ}$5866 must be quite rare since we detected only one such SB4 in our sample of 196 X-ray-selected M dwarfs. Other high-resolution spectroscopic surveys for SB's have turned up very few if any tight SB4s; e.g.~Udry et al.~(1998) surveyed 3347 G dwarfs and found none, while Torres et al.~(2006) flagged two SB4's in a X-ray-selected sample of 1511 stars with $B-V$ $>$ 0.6. Of these two, one is BD~--22$^{\circ}$5866 (independently discovered) and the other (CPD-64 4353) is a higher mass system whose orbital limits are unconstrained. 
These limit the frequencies of SB4s to 2 in 4000--5000, with only BD~--22$^{\circ}$5866 consisting of four low-mass stars. Note that Torres et al.'s and our samples are biased towards active close-in SBs with strong X-ray emission and high orbital velocities more easily resolved with the CCF, making the intrinsic frequency of these types of systems even lower. 

Though there are a couple of dozen hierarchical quadruples listed in the literature, they mostly consist of two visually resolved SB2s at wide ($d_\mathrm{AB}\gtrsim$30 AU) separations, with a handful of them exhibiting four sets of spectral lines. (See Torres et al.~2007 and references therein.) We are aware of only one previously known spatially $un$resolved quadruple-lined spectroscopic binary: XY Leo, a W UMa-type binary (Barden 1987, Pribulla et al.~2007), making BD~--22$^{\circ}$5866 only the second such system studied and the first to be comprised of only low-mass stars.\footnote{GG Tau is a famous pre-main sequence low-mass quadruple system but in a much wider orbital configuration than BD~$-$22$^{\circ}$5866. The distance between binaries A and B is 1414 AU with the Aa + Ab and Ba + Bb separated by 35 and 207 AU, respectively (White et al.~1999).} The system also contains the thirteenth published eclipsing binary (EB) with component masses less than 0.7 M$_\sun$, and the only one known to be in a quadruple system.

\section{The spectra}\label{spectra}

We acquired 4 high-resolution \'echelle spectra of BD~--22$^{\circ}$5866, 2 with the High Resolution Echelle Spectrometer (HIRES; Vogt et al.~1994) on the Keck I 10-m telescope and 2 with the \'Echelle SpectroPolarimetric Device for
the Observation of Stars (ESPaDOnS; Donati et al.~2006) on the Canada-France-Hawaii 3.6-m telescope, both located on the summit of Mauna Kea.

We used the 0.861$\arcsec$ slit with HIRES to give a spectral resolution of $\lambda$/$\Delta\lambda$$\approx$58,000. The upgraded detector consists of a mosaic of three 2048 x 4096 15-$\micron$ pixel CCDs, corresponding to a blue, green and red chip spanning 4900 -- 9300 \AA. To maximize the throughput near the peak of a M dwarf spectral energy distribution, we used the GG475 filter with the red cross-disperser. The data product of each exposure is a multiple-extension FITS file from which we reduce and extract the data from each chip separately.  

ESPaDOnS is fiber fed from the cassegrain to coud\'e focus where the fiber image is projected onto a Bowen-Walraven
slicer at the spectrograph entrance. With a 2048$\times$4608-pixel CCD detector, ESPaDOnS'
`star+sky' mode records the full spectrum over 40 grating orders covering 3700 to 10400 \AA\/ at a spectral
resolution of $\lambda$/$\Delta\lambda$$\approx$68,000. The data were reduced using {\it Libre Esprit}, a fully automated reduction package
provided for the instrument and described in detail by Donati et al.~(1997, 2007).

\begin{figure}
\epsscale{.80}
\plotone{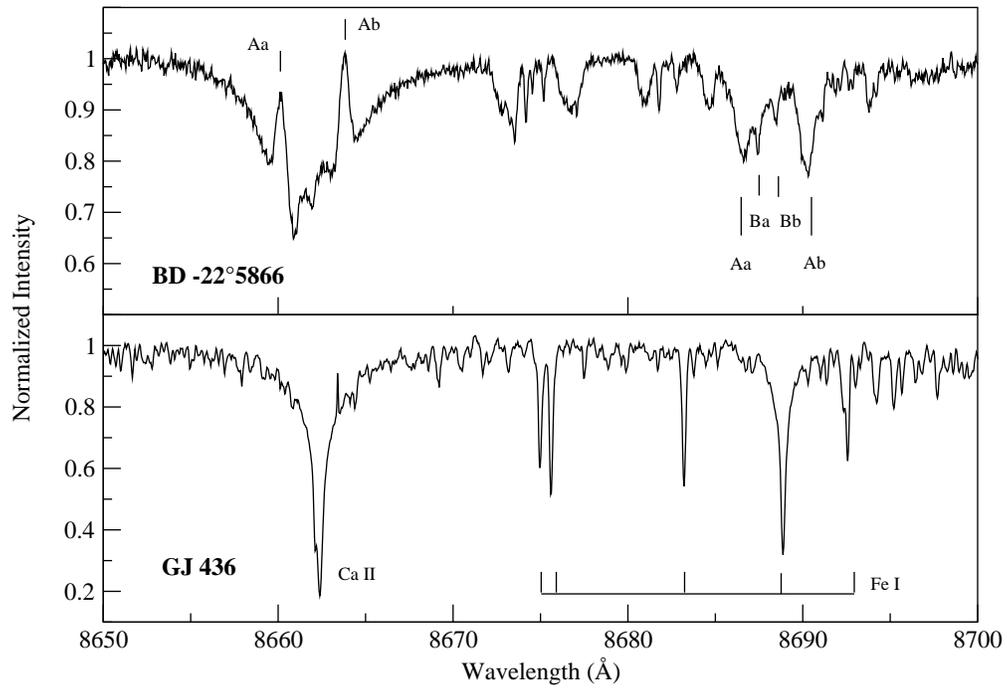}
\caption{An example spectrum of BD~--22$^{\circ}$5866 and RV standard GJ 436 observed on May 11, 2006. The spectral lines of the four components are marked by the vertical dashes. The strong Ca II emission from the Aa and Ab components is apparent at 8660 and 8664 \AA.
\label{spec}}
\end{figure}

Each stellar exposure is bias-subtracted and flat-fielded for pixel-to-pixel sensitivity variations. After optimal
extraction, the 1-D spectra are wavelength calibrated with a Th/Ar arc. Finally
the spectra are divided by a flat-field response and corrected for the heliocentric velocity.
The final spectra were of moderate S/N reaching $\approx$ 70 per pixel at 7000 \AA. Each night, spectra were
also taken of an A0V standard star for telluric line correction and an early-M radial velocity standard. A log of the observations is presented in Table~\ref{observations}, and a  portion of a spectrum is shown in Figure~\ref{spec}.

The integrated spectral type (SpT) was measured to be M0 from the ratio of the band indices of TiO~$\lambda$7140 to TiO~$\lambda$8465 defined by Slesnick et al.~(2006). This is slighter earlier than the M0.5 determined from low resolution spectra using the TiO~$\lambda$7140 index by Reid et al.~(2004).

\begin{figure}
\epsscale{.80}
\plotone{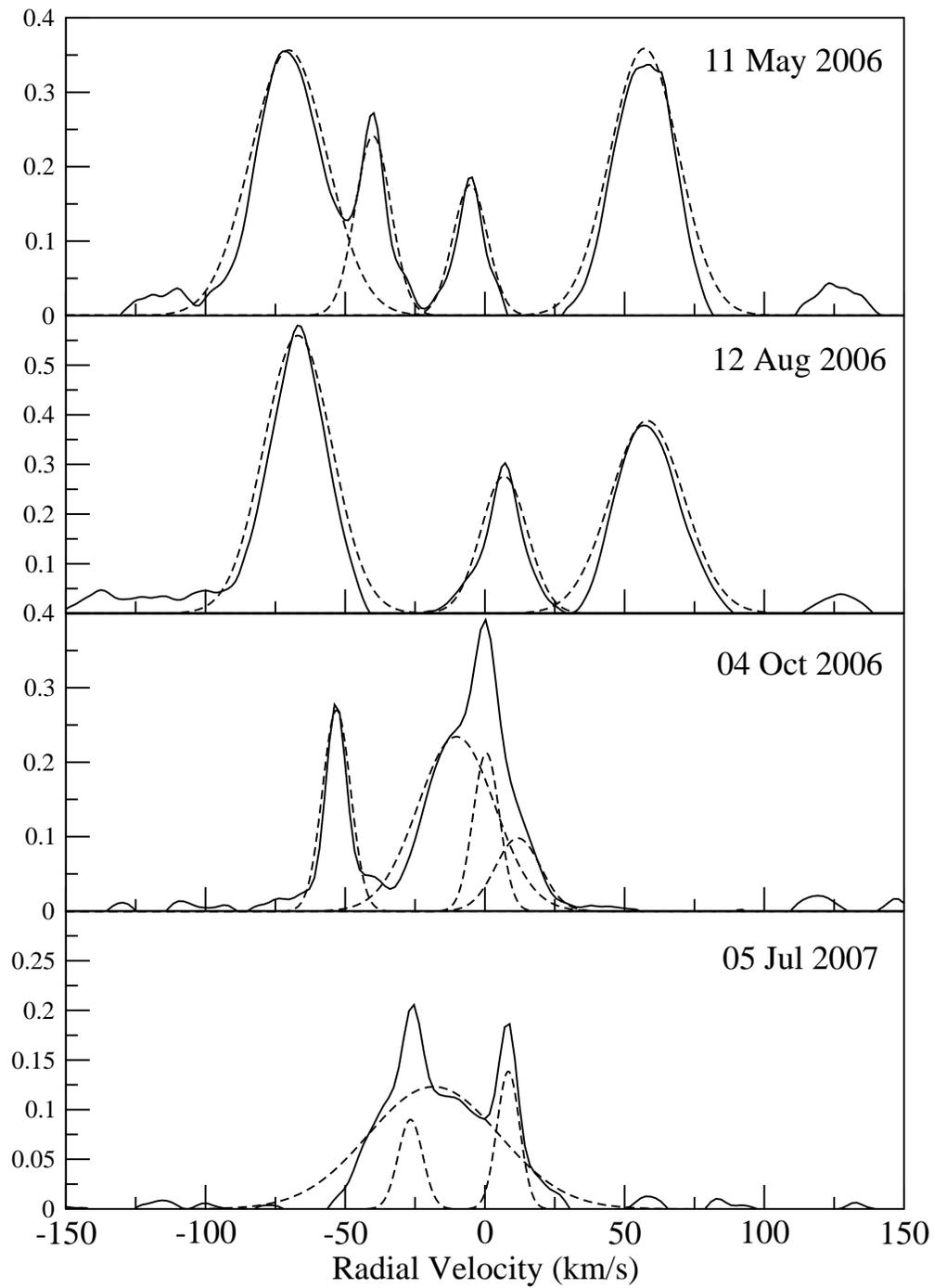}
\caption{The cross-correlation function (with baseline removed) for BD~--22$^{\circ}$5866 as measured against a RV standard star on each of the four nights.  The dashed curves are the gaussians fitted to the CCF peaks.
\label{ccf}}
\end{figure}

We cross-correlated each of 7 orders between 7000 and 9000 \AA\/ of each stellar spectrum with a RV standard of similar spectral type using IRAF's\footnote[1]{IRAF (Image Reduction and Analysis Facility) is distributed by
the National Optical Astronomy Observatories, which is operated by the Association of Universities for Research in Astronomy, Inc.~(AURA) under cooperative agreement with the National Science Foundation.} {\it fxcor}
routine (Fitzpatrick 1993). We excluded the Ca II infrared triplet (IRT)\footnote{Since the Ca II emission was clearly stronger in two of the four stars and potentially variable, it was important to eliminate the IRT from the cross-correlation.} and regions of strong telluric absorption in the cross-correlation and removed a broad, low order curvature from each CCF which varies from order to order and is due to differences in the continuum between BD~--22$^{\circ}$5866 and the RV templates. A representative cross-correlation function (CCF) for each of the four observations is shown in Figure~\ref{ccf}.

We were able to clearly resolve the CCF peaks of all four stellar components on May 11, 2006. This allowed us to estimate the spectral types of the four stars assuming a flux-weighted relation (Cruz \& Reid 2002) between component and integrated spectral types: $\mathrm{SpT_{int}} = (f_\mathrm{Aa} \mathrm{SpT_\mathrm{Aa}} + f_\mathrm{Ab} \mathrm{SpT_\mathrm{Ab}} + f_\mathrm{Ba} \mathrm{SpT_\mathrm{Ba}} + f_\mathrm{Bb} \mathrm{SpT_\mathrm{Bb}})/(f_\mathrm{Aa}+f_\mathrm{Ab}+f_\mathrm{Ba}+f_\mathrm{Bb})$, where $f_\mathrm{Aa}$, $f_\mathrm{Ab}$, $f_\mathrm{Ba}$ and $f_\mathrm{Bb}$ are derived from the integrated flux of the gaussians fits to the four cross-correlation peaks. In this case, $f_\mathrm{Aa}\approx f_\mathrm{Ab}$ and $f_\mathrm{Ba}\approx f_\mathrm{Bb}\approx 0.3f_\mathrm{Aa}$ such that the difference in \emph{I} magnitude between Aa and Ba is 1.2 mag. To measure the error associated with these relative fluxes, we cross-correlated the spectrum with three template spectra of stars with low $v$sin$i$ and differing SpTs. These were GJ~2079 (K7, $v$=4.1 km~s$^{-1}$ as calculated from $P_{rot}$ of 7.78 d; Pizzolato et al.~2003), GJ 908 (M1, $v$sin$i$=3 km~s$^{-1}$; Glebocki et al.~2000), and GJ 436 (M2.5, $v$sin$i$$\leq$1 km~s$^{-1}$; Glebocki et al.~2000). The average results give $f_\mathrm{Aa}:f_\mathrm{Ab}:f_\mathrm{Ba}:f_\mathrm{Bb}$ = 1 : 0.99 : 0.33 : 0.23. These agree to better than 15\%, with the bulk of the error lying with $f_\mathrm{Bb}$ whose integrated fluxes ranged from 0.2 to 0.3 of $f_\mathrm{Aa}$.

From the spectra alone, we estimated the component spectral types to be K7 + K7 and M1 + M2. 
These spectral types are consistent with the distinctly redder location of BD~--22$^{\circ}$5866 on a color-color diagram than expected if it were a single M0 dwarf.  For example, it lies $\sim$0.7 mags redder than expected for an M0 dwarf on a $(V-J)$-$(I-J)$ plot (Reid \& Hawley 2005).
BD~--22$^{\circ}$5866 is 3.7 times more luminous than a single M0 dwarf, and correcting for this over-luminosity, the system's photometric distance is at about 51 pc.


We measure the RVs (listed in Table~\ref{observations}) of each component from the gaussian peaks fitted to the CCFs, taking the average of all orders, with a RMS typically less than 1 km~s$^{-1}$. Severe blending of the CCF peaks resulted in poor gaussians fits to the two latest observations. One additional set of RVs was extracted from a low-S/N spectrum collected using FEROS on La Silla's 1.5-m telescope.  The instrument and data reduction of this measurement is presented in Torres et al.~2006.

\section{The light curve}\label{lc}

With the high eclipsing probability ($>$ 10\%) of the tight Aa+Ab binary, we searched for existing photometry of this system in the database of the UK Wide-Angle Search for Planets (WASP) project. The WASP project is a wide-angle, robotic photometric monitoring campaign designed to
detect significant numbers of transiting extrasolar planets using a single pass band from 4000 to 7000 \AA.
The project instrumentation, infrastructure, and data processing are
are described in detail in Pollacco et al.~(2006).  In short, SuperWASP (La Palma, Canary Islands, Spain) and WASP-South (Sutherland, South Africa)
instruments each consist of 8 small aperture camera lenses (8 $\times$ 200mm, f/1.8)
backed by wide-format CCDs with a plate scale of $14.2^{\prime\prime}$/pixel.
The cameras are attached to a single robotic mount such that the field-of-view
of a single WASP pointing is approximately $15^{\circ}\times 30^{\circ}$.
The mount scans the sky taking repeated individual exposures in a single
declination band over the observable sky every clear night to obtain a time-series
of images with typical sampling rate of $\sim 8$~minutes.   The raw images are processed via a
custom-built pipeline with aperture photometry performed using a 3.5~pixel aperture (48$^{\prime\prime}$) on all observed
stars.
The resulting differential photometry
light curves are searched for periodic dips in brightness using a modified
box-least squares search algorithm which is outlined in Collier Cameron et al.~(2006).

\begin{figure}
\epsscale{.80}
\plotone{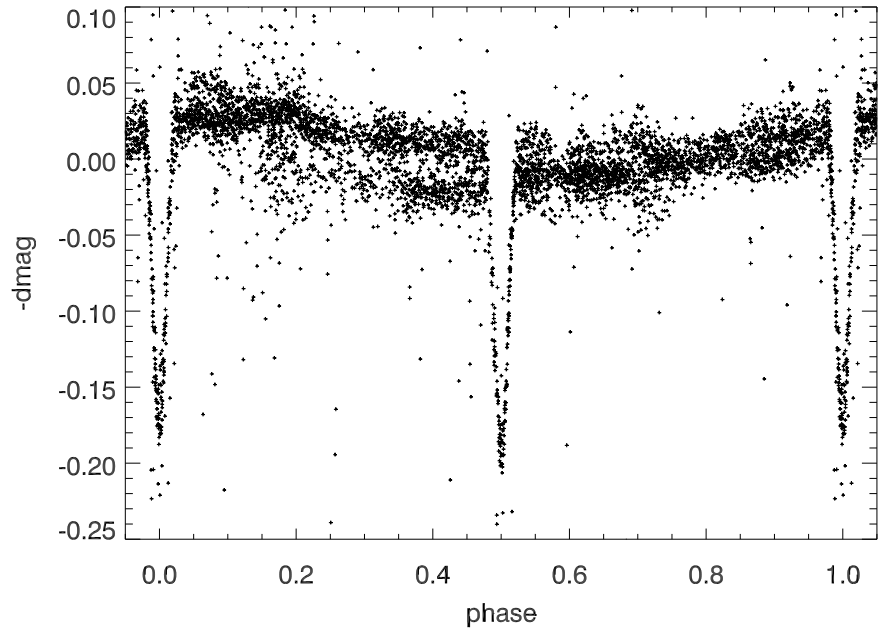}
\caption{The raw WASP-S light curve of BD~--22$^{\circ}$5866 phased on the eclipse period of 2.21 days. The difference in starspot intensity from the two epochs is clear in the double-lined region of the light curve.
\label{raw_lc}}
\end{figure}

\begin{figure}
\epsscale{.80}
\plotone{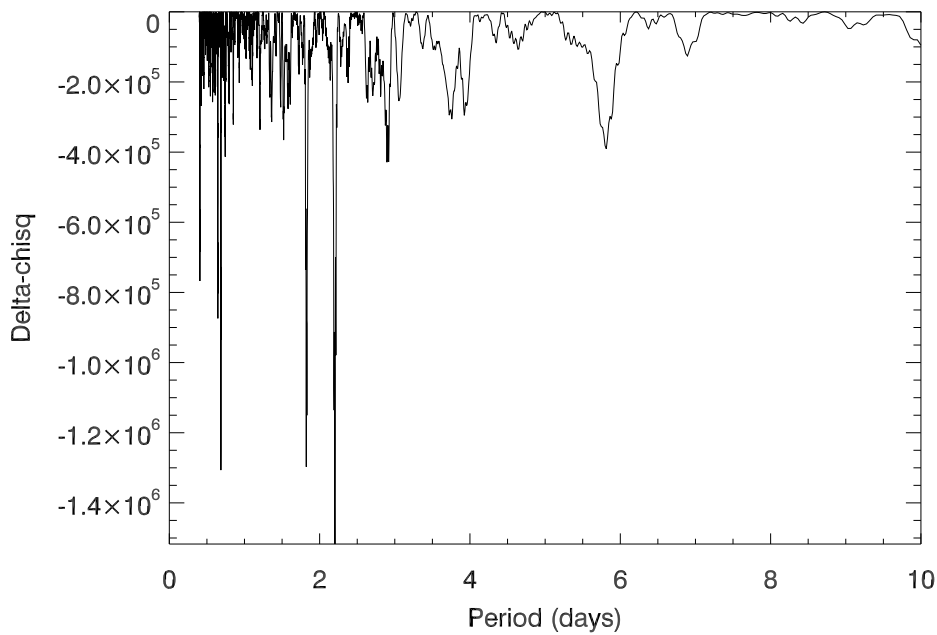}
\caption{Periodogram of the sine-like out-of-eclipse variability in the WASP-S light curve. We tested 5000 frequencies between 0.4--20 days and found the parameters of a sine-curve (amplitude, phase and zero-point offset) that best fit the out-of-eclipse variability at each trial period. The peak at $\approx$2.2 days is clear which is nearly identical to the orbital period of the binary, thus indicating the stars are tidally locked.
\label{pdm}}
\end{figure}

The high-cadence, high precision time-series photometry produced for millions of
bright stars as part the WASP transit survey is extremely good for detecting
and characterizing eclipsing binaries in addition to the
primary science goal of transiting planet detection.  
The WASP-South data set on BD~--22$^{\circ}$5866 consists of 9531 photometric measurements obtained over two observing seasons (May 2006 -- October 2007).  The observed light curve shows periodic ($P$=2.21~days), nearly equal depth ($\approx 0.2$~mag)  primary and secondary eclipses identifying the tight K7 + K7 pair as an eclipsing binary. The raw WASP light curve is shown in Figure~\ref{raw_lc} phased according to the eclipse ephemeris (see Table~\ref{orbpars}).  In addition to the eclipses, the photometry also exhibits modulating out-of-eclipse variability with an amplitude of 0.01--0.03~mags which is likely caused by asymmetric starspots on the surface of the individual binary components. The measured period of this variability as determined from the strongest peak of the periodogram (Figure~\ref{pdm}) is nearly identical to the binary orbital period and tracks the rotation of the star indicating the system is tidally synchronized. This is expected since the synchronization time-scale for the short-period binary A is only 70 Myr (Zahn 1977), several orders of magnitude shorter than for binary B, making it the likely source of the observed X-ray and UV flux. The RVs also confirm that Aa and Ab are the sources of the moderate emission we see in the lines of the Ca II IRT (Figure~\ref{spec}) and H$\alpha$ (EW$_\mathrm{H\alpha}$(Aa)=--0.37 $\pm$ 0.01 \AA, EW$_\mathrm{H\alpha}$(Ab)=--0.55 $\pm$ 0.01 \AA). We therefore conclude the system is not particularly young, consistent with the fact that we see no sign of Li ($\lambda$6708\AA, EW$_\mathrm{Li}$ $<$ 21 m\AA). Comparing these characteristics of binary A with the late-type systems of the Strassmeier et al.~2003 catalog of chromospherically active binary stars, we can classify BD~--22$^{\circ}$5866~A as BY Dra binary.

\begin{figure}
\epsscale{.80}
\plotone{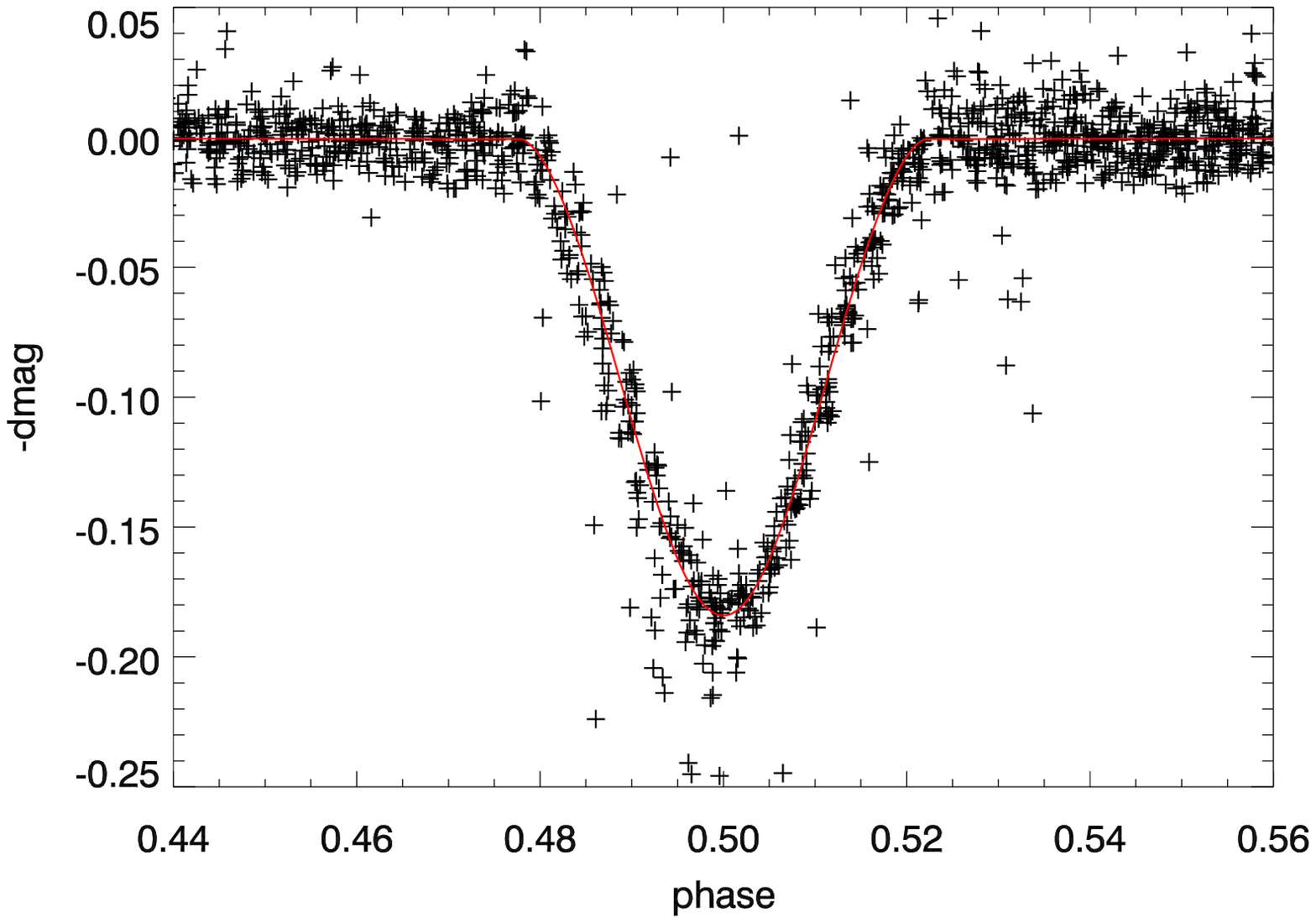}
\caption{Five radial velocities measured for Aa (circles) and Ab (squares) of BD~--22$^{\circ}$5866. The best fit sine curves assume $e=0$ and equal velocity amplitudes and is fit only to the Mauna Kea data (white points) outside of eclipse. The La Silla data are plotted as black points and display a significant shift in $\gamma$ due to the orbital motion of the two binaries. Measurement errors in both RV and phase are within the size of the points. The fits give $K_{\mathrm{Aa}}=K_{\mathrm{Ab}}=86.3$ km~s$^{-1}$ and $\gamma=-5.1$ km~s$^{-1}$.
\label{rv}}
\end{figure}

Due to the eclipsing nature of the K7 + K7 binary, the individual stellar radii, temperature ratio and orbital inclination can be measured directly from the light curve.  We first remove the out-of-eclipse variability before modeling the eclipsing system and deriving the physical properties of the individual stars.  Because the amplitude of the modulation is changing slowly over the course of the observations, we separate the photometric data into 3-week time bins and remove the variability of each segment independently.  For each time segment, we solve for the coefficients of a sine curve that best fits the out-of-eclipse data using least-squares minimization and then subtract this model from the observed photometry. We then derive the properties of the system using an implementation of the EBOP eclipsing binary modeling code (Southworth et al.~2004, Popper \& Etzel 1981).   The rectified light curve around the primary and secondary eclipses is shown in Figure~\ref{lc} with the best fitting model over-plotted.  The results of the analysis give individual stellar radii of $R_{Aa} = 0.614\pm 0.045 R_{\sun}$ and $R_{Ab} = 0.598\pm 0.045 R_{\sun}$ (using the semi-major axis\footnote{We have used the luminosity ratio of stellar components Aa and Ab derived from the cross-correlation functions (See Section~\ref{spectra}) as an independent constraint on the ratio of the radii and the amplitude of the radial velocity curve described in the following section to derive the orbital separation.} derived from the RV curve, see Figure~\ref{rv} and Table~\ref{orbpars}), an orbital inclination of $85.5^{\circ} \pm 1^{\circ}$, a temperature ratio $T_{Ab}/T_{Aa} = 0.98 \pm 0.02$, and an eccentricity of 0.00 $\pm 0.01$.





\section{The orbital configuration}


Although there are not many RV measurements, we can determine well-constrained masses due to the accurate ephemeris derived from the photometry. With the precise eclipse period, we are able to plot our RV measurements to determine the remaining orbital parameters for the K7 + K7 binary.  To reduce the number of fitted parameters needed to derive the component masses,  we adopt a mass ratio of 1.0 for the binary because the two components have nearly the same flux and temperature values. We solved for a single RV amplitude and systemic velocity $\gamma$ which best fit the Keck and CFHT RVs outside of eclipse. The 5-year span between the Mauna Kea and the La Silla observations is a significant fraction of the A + B orbital period, and the corresponding shift in $\gamma$ due to the presence of binary B is apparent in the offset of those points in Figure~\ref{rv}. The true systemic velocity of the entire system is close to --9 km~s$^{-1}$ measured by taking the average all observed RVs.

The mass for each K7 dwarf derived from the RV curve is 0.59 M${_\sun}$. We estimate the uncertainty on this mass to be $\approx$5\% based on the potential deviations from a mass ratio of 1. Additional RV measurements will allow us to derive component masses with $<$1\% accuracy. The relative flux levels of the CCF peaks indicate that these two stars are in between a K7 and M0 dwarf, consistent with their derived masses. The system parameters are summarized in Table~\ref{orbpars}.

\begin{figure}[htb]
\begin{center}
\includegraphics[angle=0, width=5in]{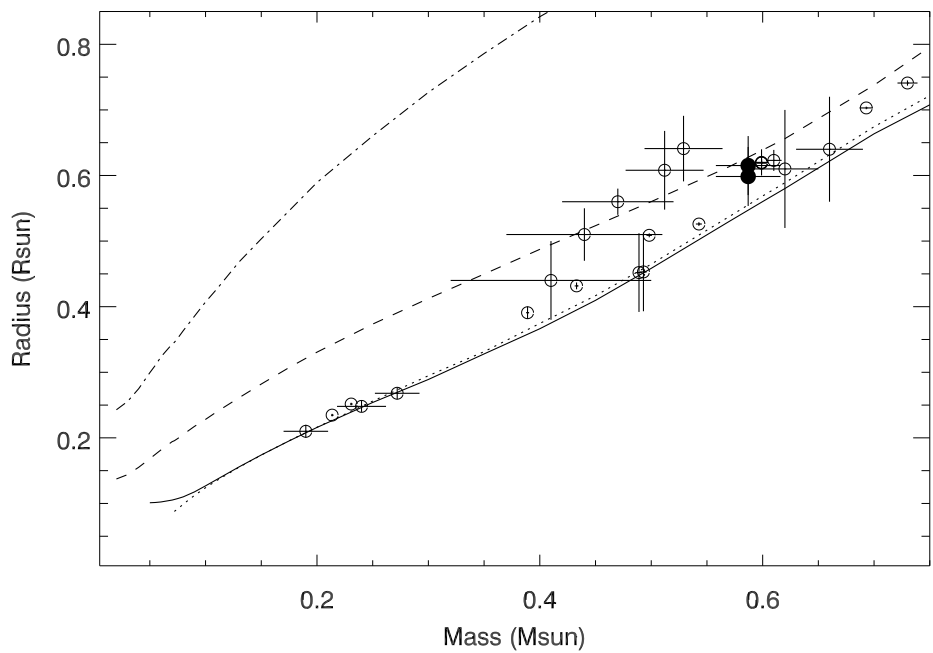}
\caption[]{Mass versus radius for known main sequence eclipsing binaries with masses, M$<0.7$~M$_{\odot}$.  The new M-dwarf EB components, BD~--22$^{\circ}$5866 Aa and Ab are shown as solid circles.  The literature data on known M dwarf-M dwarf EBs (open circles) are taken from the references listed in Table~\ref{eb}. 
The lines show the theoretical, solar metallicity mass-radius relation with ages of 10 Myr (dot-dashed), 50 Myr (dashed), 500 Myr (dotted), and 5 Gyr (solid) (Baraffe et al.~1998).
\label{mr}}
\end{center}
\end{figure}

This pair is only the thirteenth confirmed EB with component masses less than 0.7 M$_{\sun}$. Having model-independent masses and radii for both components is an essential part of testing stellar evolution models (Baraffe et al.~1998). Although the measured properties of the K7 binary components are consistent with being on the main sequence, the two stars appear $\approx$10\% larger in radii for their masses than predicted by the stellar evolution models (Figure~\ref{mr}), in agreement with recent measurements of the other low-mass EBs  with 10-20\% larger radii than expected. (See Table~\ref{eb})  Some empirical data
suggest this is due to the effects of magnetic activity
(L\'opez-Morales 2007) while other data show no
correlation with activity, and instead find metallicity
to be the dominant cause (Berger et al. 2006).


With regards to the rest of the quadruple system, the maximum velocity separation measured for the lower-mass pair (Ba + Bb) is 52 $\pm$ 0.8 km~s$^{-1}$. This coupled with the estimated spectral types (and masses 0.49 and 0.44 M$_{\sun}$ for a M1 and M2 dwarf, respectively; Reid \& Hawley 2005), allows us to set upper limits on its orbital period and semi-major axis of 62 days and 0.30 AU. Though there is no indication of eclipses for this wider binary, we cannot determine if Ba and Bb are in a coplanar orbit with binary A.  With a grazing eclipse angle of $>$89$^{\circ}$, Ba + Bb would not exhibit eclipses even if coplanar with Aa + Ab.   

On 11 May, 2006, we observed a maximum relative orbital velocity of 17.5 $\pm$ 2.3 km~s$^{-1}$ for the A and B
binaries. If we treat the system as two single objects in Keplerian orbit, then this velocity measurement sets an upper
limit of $a_\mathrm{AB}$sin$i_\mathrm{AB}$$ \leq 6.1$ AU and $P_\mathrm{AB} \leq 3764$ d. At this separation, the binary pair A + B is resolvable with ground-based adaptive optics imaging.\footnote{Daemgen et al.~(2007) observed BD~--22$^{\circ}$5866 to have no physical companions within 0.08$\arcsec$ using adaptive optics, setting an upper limit on the physical separation between A and B of $d_\mathrm{AB}<4.1$ AU at 51 pc from Earth at the observed orbital phase.} Though the change in $\gamma$ of $\approx$17 km~s$^{-1}$ in the five years between observations implies that the period is likely closer to 6 years (and then $a_\mathrm{AB}$sin$i_\mathrm{AB} \sim$ 4 AU) rather than the 10 years of the maximal orbit. 

\begin{figure}[htb]
\begin{center}
\includegraphics[angle=270, width=5in]{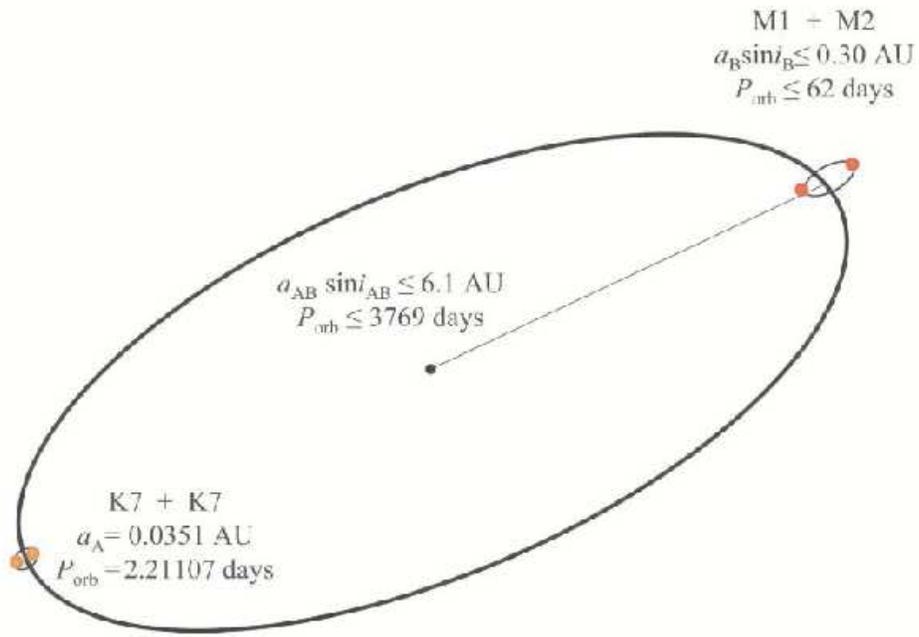}
\caption[]{The maximal orbit of the low-mass quadruple-lined SB, BD~--22$^{\circ}$5866 drawn nearly to scale. 
\label{orbit}}
\end{center}
\end{figure}

A schematic of this ESB4 with its limiting orbital parameters is shown in Figure~\ref{orbit}. In order to use the empirical criterion for the orbital stability of binary systems (Eggleton \& Kiseleva 1995), we approximate this system to be a hierarchical triple by treating the tight Aa+Ab binary as a single mass, and conclude that even for high values of eccentricity, BD~--22$^{\circ}$5866 is very stable.


\section{A need for a primordial circumquadruple disk?}

In order to reproduce the typical observed separations of $\sim$40 AU in binary systems, Sterzik et al.~(2003) determined that a multiple star system must undergo a phase of dynamical evolution after the fragmentation stemming from isothermal collapse of the molecular cloud. However, the formation of very close binaries with $a$ on the order of 1 AU, as is the case for BD~--22$^{\circ}$5866~AB, is still somewhat unclear. Tokovinin et al.'s (2006) determination that the vast majority of short-period binaries must have tertiary components with which angular momentum can be exchanged may explain the tight orbit of Aa+Ab, yet also implies that $a_\mathrm{AB}$ must have been even smaller in the past. This same argument may imply that a distant fifth component to the system might have interacted with binaries A and B to bring them to their small separation. However, no common proper motion companions were found within 2000 AU of the system.\footnote{We searched the Naval Observatory Merged Astrometric Dataset (NOMAD; Zacharias et al.~2005) which compiles data from several catalogs including the 2MASS catalog (Skrutskie et al.~2006). With a K-band magnitude limit of 16, $M_K$ is limited to 12.4 mags at the system distance of 51 pc. This sets the maximum mass of any potential companion to 20--50 M$_J$, the mass of an old field mid- to late-L dwarf.} We speculate that an earlier dynamical process reduced the physical size of the system, such as binary-disk interactions within a circumquadruple disk. 

This is supported by the simulations by Bate et al.~(2002) of binary- and triple-star formation which predict the shrinking of orbital distances to $\lesssim$10 AU through accretion and interactions with circumbinary and circumtriple gas-rich disks on the time-scale of less than 10$^5$ years. Though in the case of a hierarchical triple, the tertiary component is always in a wide orbit of more than 30 AU from the tight binary.  They also calculate that the specific angular momentum exchange will drive the masses of the two components towards equality which is marginally the case for the A and B binaries of BD~--22$^{\circ}$5866 which has a mass ratio $q \approx 0.7$.\footnote{This may also explain the even tighter, higher mass system of VW LMi consisting of one non-eclipsing detached binary and one eclipsing contact binary at a separation of $d_\mathrm{AB}=1.24$ AU and with solar-mass components (Pribulla et al.~2006).} 

There is no theoretical discussion in the literature regarding circumquadruple disks around low-mass stars. Yet it is reasonable to assume that  BD~--22$^{\circ}$5866 must have once had such a gas-rich primordial disk, which likely lasted for less than 10$^5$--10$^6$ years (Artymowicz \& Lubow 1994). This would imply that the components accumulated their masses {\it and} interacted dynamically with each other (Sterzik et al.~2005) and their disk during the same early stages of their evolution. A systematic interferometric investigation of young pre-main sequence stars to search for additional low-mass quadruple systems will help test our hypothesis of stellar interactions with a circumquadruple disk, since these very young systems should be wider.

\section{Future observations}\label{summary}


Since the BD~--22$^{\circ}$5866 system is both bright and contains short orbital periods, we encourage those actively monitoring binaries to include this low-mass system in their programs. Spectroscopic monitoring will yield the component velocities for which specially designed techniques, such as the broadening function formulism of Rucinski et al.~(2002) or the four-dimensional cross-correlation method by Torres et al.~(2007), are well suited. These are necessary to determine the Keplerian orbital parameters, which will provide more accurate mass ratios, and if the two remaining inclinations can be measured, the masses of the individual components, arguably the most important stellar property in the context stellar evolution (e.g., interferometric observations (i.e.~Boden et al.~2005) offer very high spatial resolution and could measure the separation and inclination of the A+B system.) In addition, a parallax measurement to determine a more accurate distance to the system would be very beneficial.

With the potential of these observations to yield all four stellar masses as well as the three orbital inclinations, the BD~--22$^{\circ}$5866 system could provide key constraints to possible formation scenarios, as well as the ability to test the hypothesis that orbital evolution within a disk would create coplanar orbits (Bonnell \& Bate 1994). Clearly, intensive follow-up observations of this rare eclipsing SB4 are warranted.


\acknowledgements

E.S appreciates useful discussion with George Herbig, Slavek Rucinski, Katelyn Allers and Nader Haghighipour, as well as helpful suggestions from the anonymous referee. Also, thank you to Nick Dunstone for
his quick and independent discovery of the eclipses in the WASP database and the CFHT and Keck staff for their care in setting up the instruments
and support in the control room. Research funding from the NASA Postdoctoral Program (formerly the NRC Research Associateship) for E.S. is
gratefully acknowledged. This material is based upon work supported by the National Aeronautics and Space Administration through the NASA
Astrobiology Institute and the NASA/GALEX grant program under Cooperative Agreement Nos. NNA04CC08A and NNX07AJ43G issued through the Office of
Space Science. M.C.L. acknowledges support from the Alfred P. Sloan Research Fellowship.


\clearpage
\begin{deluxetable}{ccllllccc}\label{observations}

\tabletypesize{\footnotesize}
\tablecaption{Observations \& Radial Velocities\label{observations}}
\tablewidth{0pt}
\tablehead{
\colhead{Date} &
\colhead{Instrument} &
\colhead{HJD-2450000} &
\colhead{$\phi_{\mathrm{Aa+Ab}}$} &
\colhead{RV\tablenotemark{a} of Aa, Ab, Ba, Bb (km s$^{-1}$)} &
\colhead{RV STD, SpT\tablenotemark{b}} &
}

\startdata

11-May-06	&	Keck I/HIRES	&	3867.12294	&	0.12994 &	$-69.0, +58.4, -40.2, -5.4$	&	GJ 436\tablenotemark{c}, M2.5		\\
12-Aug-06	&	Keck I/HIRES	&	3959.97821	&	0.12549 &	$-66.9, +57.3, {\it+7.5, +7.5}$	&	GJ 908, M1		\\
4-Oct-06	&	CFHT/ESPaDOnS	&	4013.83591	&	0.48365 &	${\it-10.2,+0.5},-53.2, +0.3$	&	GJ 205, M1.5		\\
5-Jul-07	&	CFHT/ESPaDOnS	&	4287.99493	&	0.47725 &	${\it -18.0,-18.0},-25.9, +8.5$	&	GJ 821, M1		\\
\\											
31-Aug-01\tablenotemark{d}	&	FEROS/1.5m	&	2152.80326	&	0.79647 &	$+64.5, -109.1, 3.7, -15.0$	&	$\tau$ Ceti, G8.5\\
\enddata

\tablenotetext{a}{For the Keck and CFHT data, heliocentric RVs are determined by taking the average of the RVs measured individually from 7 spectral orders with RMS $\lesssim $1 km s$^{-1}$. Values in italics represent those from CCF peaks that are strongly blended.} 

\tablenotetext{b}{The RVs for these standards are published by Marcy \& Benitz (1989), except GJ 821, which is from Nidever et al.~2002.}

\tablenotetext{c}{The velocity amplitude of the Neptune-mass planet orbiting GJ 436 is only 0.019 km~s$^{-1}$ (Butler et al.~2006), negligible for our purposes.}

\tablenotetext{d}{Details of this spectrum can be found in Torres et al.~2006.}

\end{deluxetable}

\begin{deluxetable}{lcccc}\label{orbpars}

\tabletypesize{\footnotesize}
\tablecaption{Parameters for Binary Aa + Ab\label{orbpars}}
\tablewidth{0pt}
\tablehead{
\colhead{Quantity} &
\colhead{Value} &
\colhead{Error} &
}

\startdata


Period (days)	&	2.21107(5)	&	0.000004	\\
Epoch (HJD $-$ 2450000)	&	3937.5900(0)	&	0.00041	\\
$i$ (degrees)	&	85.5	&	1.0	\\
$e$				&	0.00		&	0.01\\
$\omega$ &	82\tablenotemark{a}	&	--	\\	
$R_\mathrm{Aa}/a$	&	0.0814	& 0.0060		\\
$R_\mathrm{Ab}/a$	&	0.0792	& 0.0058		\\
$T_{eff,\mathrm{Aa}}/T_{eff,\mathrm{Ab}}$	&	0.98	&	0.02 	\\
$M_\mathrm{Aa}$, $M_\mathrm{Ab}$ (M$_{\sun}$)	&	0.5881	&	0.029	\\
$a$ (AU)	&	0.0351	&	0.0024	\\
$R_\mathrm{Aa}$ (R$_{\sun}$)	&	0.614	&	0.045	\\
$R_\mathrm{Aa}$ (R$_{\sun}$)	&	0.598	&	0.045	\\

\enddata

\tablenotetext{a}{With an eccentricity of 0, the angle of periastron $\omega$ is very poorly constrained.}

\end{deluxetable}

\begin{deluxetable}{llllllccc}\label{eb}

\tabletypesize{\footnotesize}
\tablecaption{Known low-mass EBs\label{eb}}
\tablewidth{0pt}
\tablehead{
\colhead{Name} &
\colhead{Mass} &
\colhead{$\delta$M} &
\colhead{$R$} &
\colhead{$\delta$R} &
\colhead{Ref.} &
\\
\colhead{} &
\colhead{M$_{\sun}$} &
\colhead{M$_{\sun}$} &
\colhead{R$_{\sun}$} &
\colhead{R$_{\sun}$} &
\colhead{} &

}

\startdata

RXJ0239.1 A	&	0.730	&	0.009	&	0.741	&	0.004	&	L\'opez-Morales \&	 Shaw 2007 \\
RXJ0239.1 B	&	0.693	&	0.006	&	0.703	&	0.002	&		\\
2MASS J01542930+0053266 A	&	0.66	&	0.03	&	0.64	&	0.08	&	Becker et al.~2008	\\
2MASS J01542930+0053266 B	&	0.62	&	0.03	&	0.61	&	0.09	&		\\
GU Boo A	&	0.610	&	0.007	&	0.623	&	0.016	&	L\'opez-Morales \&	 Ribas 2005\\
GU Boo B	&	0.599	&	0.006	&	0.62	&	0.02	&		\\
YY Gem A	&	0.599	&	0.0047	&	0.6191	&	0.0057	&	Delfosse et al.~1999	\\
YY Gem B	&	0.599	&	0.0047	&	0.6191	&	0.0057	&		\\
BD~--22$^{\circ}$5866 Aa	&	0.588	&	0.029	&	0.614	&	0.045	&	this work	\\
BD~--22$^{\circ}$5866 Ab	&	0.588	&	0.029	&	0.598	&	0.045	&		\\
NSVS01031772 A	&	0.5428	&	0.0027	&	0.526	&	0.0028	&	L\'opez-Morales 2006	\\
NSVS01031772 B	&	0.4982	&	0.0025	&	0.5088	&	0.003	&		\\
UNSW-TR-2 A	&	0.529	&	0.035	&	0.641	&	0.05	&	Young et al.~2006	\\
UNSW-TR-2 B	&	0.512	&	0.035	&	0.608	&	0.06	&		\\
Tres-Her0-07621 A	&	0.493	&	0.003	&	0.453	&	0.06	&	Creevey 2005	\\
Tres-Her0-07621 B	&	0.489	&	0.003	&	0.452	&	0.06	&		\\
2MASS 04463285+1901432 A	&	0.47	&	0.05	&	0.56	&	0.02	&	Hebb et al.~2006	\\
2MASS 04463285+1901432 B	&	0.19	&	0.02	&	0.21	&	0.01	&		\\
OGLE BW3 V38 A	&	0.44	&	0.07	&	0.51	&	0.04	&	Maceroni \&	 Montalb\'an 2004\\
OGLE BW3 V38 B	&	0.41	&	0.09	&	0.44	&	0.06	&		\\
CU Cnc A	&	0.433	&	0.0017	&	0.4317	&	0.0052	&	Delfosse et al.~1999	\\
CU Cnc B	&	0.389	&	0.0014	&	0.3908	&	0.0094	&		\\
SDSS-MEB-1 A	&	0.272	&	0.02	&	0.268	&	0.009	&	Blake et al.~2007	\\
SDSS-MEB-1 B	&	0.240	&	0.022	&	0.248	&	0.0084	&		\\
CM Dra A	&	0.2307	&	0.001	&	0.2516	&	0.002	&	Metcalf et al.~1996	\\
CM Dra B	&	0.2136	&	0.001	&	0.2347	&	0.0019	&		\\

\enddata

\end{deluxetable}

\clearpage

\begin{figure}
\epsscale{1.0}
\plottwo{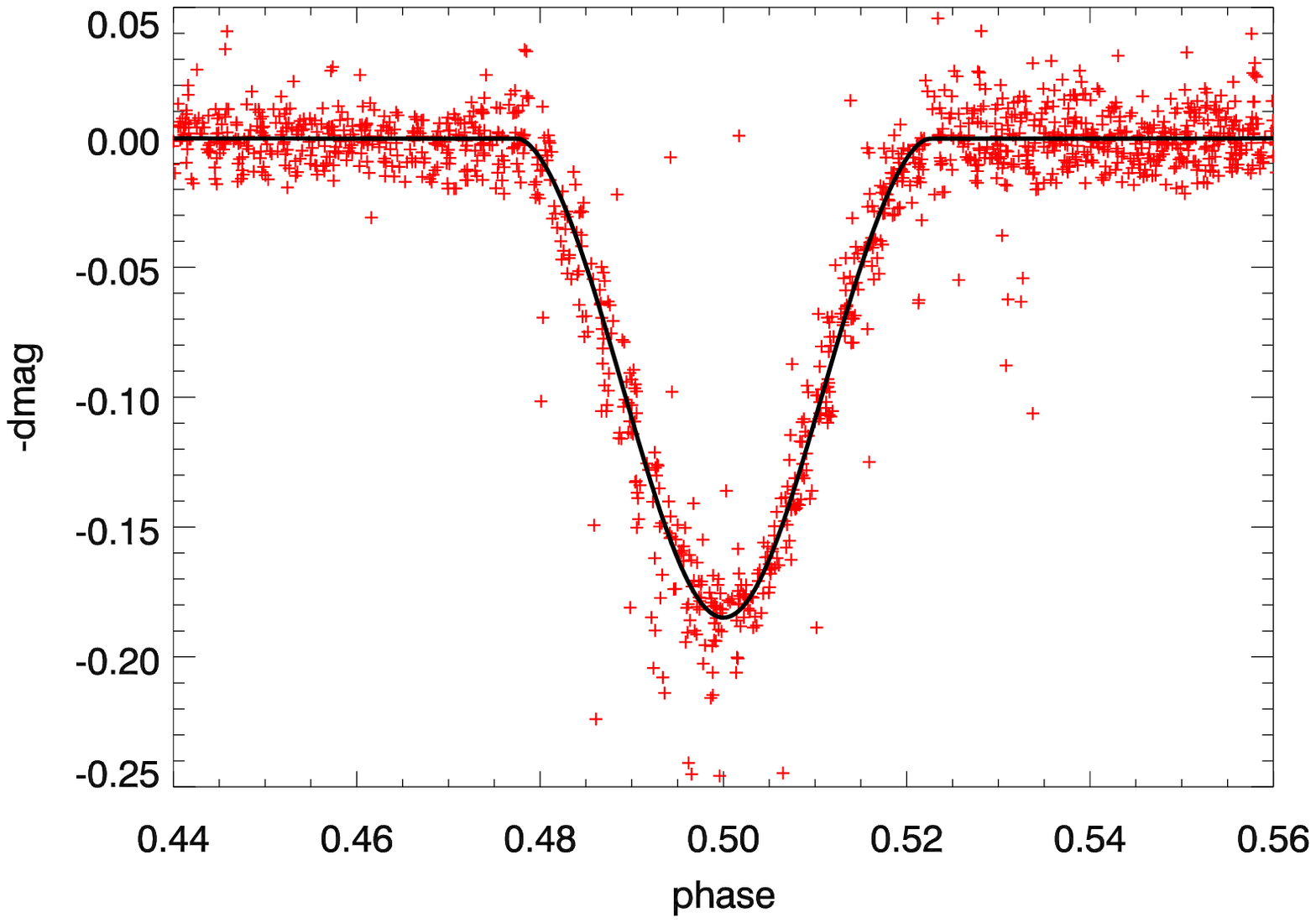}{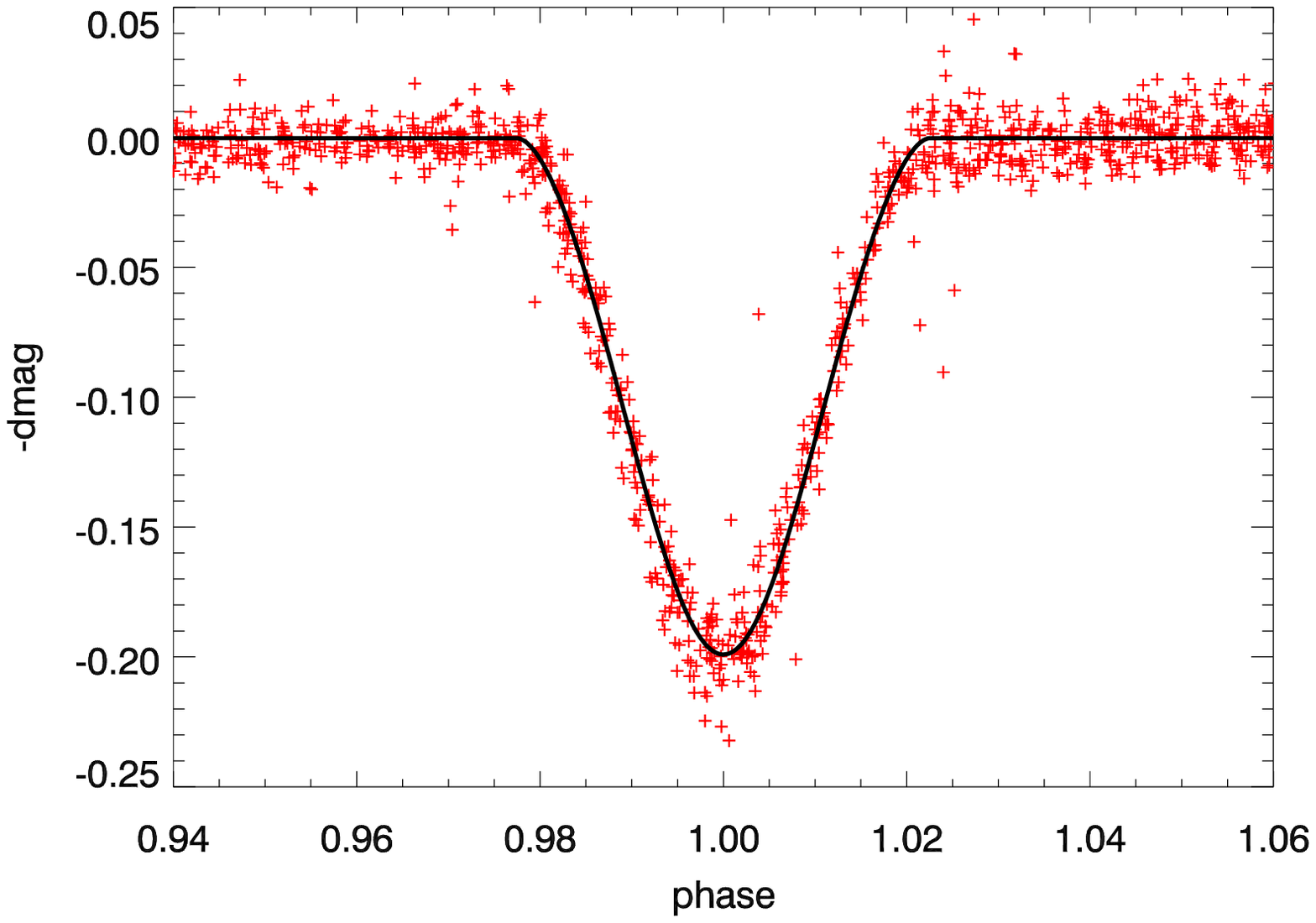}
\caption{The rectified light curves for the Aa (left) and Ab (right) eclipses of BD~--22$^{\circ}$5866 with the best model fit.
\label{lc}}
\end{figure}

\end{document}